\documentstyle[psfig]{elsart}
\begin{document}
\begin{frontmatter}

\title{Effects of Disorder in \protect\( Mg_{1-x}\protect \) \protect\(
Ta_{x}\protect \) \protect\( B_{2}\protect \) Alloys using
Coherent-Potential Approximation}

\author{P. Jiji Thomas Joseph and Prabhakar P. Singh}

\address{{\large Department of Physics, Indian Institute of Technology,
Powai, Mumbai- 400076, India}\large }

\begin{abstract}
Using Korringa-Kohn-Rostoker coherent-potential approximation in the
atomic-sphere approximation (KKR-ASA CPA) method for taking into account
the effects of disorder, Gaspari-Gyorffy formalism for calculating the
electron-phonon coupling constant \( \lambda \), and Allen-Dynes equation
for calculating \( T_{c}, \) we have studied the variation of \( T_{c} \)
in \( Mg_{1-x}Ta_{x}B_{2} \) alloys as a function of \( Ta \)
concentration. Our results show that the \( T_{c} \) decreases with the
addition of \( Ta \) for upto \( 40\, at\% \) and remains essentially
zero from \( 60\, at\% \) to \( 80\, at\% \) of \( Ta. \) We also find \(
TaB_{2} \) to be superconducting, albeit at a lower temperature. Our
analysis shows that the variation in \( T_{c} \) in \( Mg_{1-x}Ta_{x}B_{2}
\) is mostly dictated by the changes in the \( B\, p \) density of states
with the addition of \( Ta. \). 
\end{abstract}
\end{frontmatter}

\paragraph{\textmd{The experimental \cite{nag,budko1,kang,buzea} and
theoretical \cite{ann,hirsch1,belash,imada,hirsch2,budko2,kong,pps1} efforts aimed at understanding the nature of
superconductivity in \protect\( MgB_{2}\protect \) have made substantial
progress since the discovery of superconductivity in it \cite{nag}. With
an enhanced understanding of superconductivity in \protect\(
MgB_{2}\protect \), attempts are now being made to understand the changes
in the electronic structure and the superconducting properties of
\protect\( MgB_{2}\protect \) upon alloying with various elements. Such
efforts provide an opportunity to explore the possibility of obtaining
\protect\( MgB_{2}\protect \) alloys with improved superconducting
properties. }}

\paragraph{\textmd{The changes in the superconducting properties of
\protect\( MgB_{2}\protect \) due to substitution of various elements such
as \protect\( Be,\, Li,\, C,\, Al,\, Na,\, Zn,\, Zr\, \protect \) and
others have been studied experimentally \cite{buzea,felner,slusky,kazakov,suzuki,zhao,morotomo}. Some
theoretical work has also been reported within the rigid-band model. The
main effects of alloying are seen to be (i) a decrease in transition
temperature, \protect\( T_{c}\protect \), with increasing concentration of
the alloying elements except for \protect\( Zn\protect \) (and possibly
\protect\( Li\protect \)) where \protect\( T_{c}\protect \) is seen to
increase somewhat, (ii)  changes in lattice parameters \protect\(
a\protect \) and \protect\( c,\protect \) and (iii) a change in crystal
structure.}}

\paragraph*{\textmd{With a view to understand the changes in the
electronic structure and the superconducting properties as well as
to clarify the reported superconductivity in \protect\( TaB\protect
\)\protect\( _{2}\protect \) \cite{kac,laya}, we have carried out an
{\it ab initio} study of \protect\( Mg\protect
\)\protect\( _{1-x}\protect \)\protect\( Ta\protect \)\protect\(
_{x}\protect \)\protect\( B\protect \)\protect\( _{2}\protect \) alloys.
We have used Korringa-Kohn-Rostoker coherent-potential approximation
\cite{faulkner} within the atomic-sphere approximation (KKR-ASA CPA) method
for taking into account the effects of disorder, Gaspari-Gyorffy formalism
for calculating the electron-phonon coupling constant \protect\( \lambda
\protect \), and Allen-Dynes equation for calculating \protect\(
T_{c}\protect \) in \protect\( Mg_{1-x}Ta_{x}B_{2}\protect \) alloys as a
function of \protect\( Ta\protect \) concentration}. \textmd{Such an
attempt allows us to examine the possibility of superconductivity in
\protect\( TaB\protect \)\protect\( _{2}\protect \), given the
superconductivity in \protect\( MgB\protect \)\protect\( _{2}\protect \).
We have analyzed our results in terms of the changes in the total density
of states (DOS), in particular the changes in the \protect\( B\protect \)
\protect\( p\protect \) contribution to the total DOS, as a function of
\protect\( Ta\protect \) concentration.}}

\paragraph{\textmd{Based on our calculations, described below, we find
that in \protect\( Mg_{1-x}Ta_{x}B_{2}\protect \) alloys (i) the
\protect\( T_{c}\protect \) decreases with the addition of \protect\(
Ta\protect \) for upto \protect\( 40\, at\%\protect \), remains
essentially zero from \protect\( 60\, at\%\protect \) to \protect\( 80\,
at\%\protect \), and then rises to \protect\( \sim 1.8\, K\protect \) for
\protect\( TaB_{2},\protect \) (ii) the \protect\( T_{c}\protect \) for
\protect\( TaB_{2}\protect \) is much lower than reported earlier
\cite{kac}, and (iii) the variation in \protect\( T_{c}\protect \) is
mostly dictated by the changes in the \protect\( B\, p\protect \)
densities of states as more and more \protect\( Ta\protect \) are added. 
Before we describe our results, we
outline  some of the computational details.}}

\paragraph{\textmd{The charge self-consistent electronic structure of
\protect\( Mg\protect \)\protect\( _{1-x}\protect \)\protect\( Ta\protect
\)\protect\( _{x}\protect \)\protect\( B_{2}\protect \) alloys as a
function of \protect\( x\protect \) has been calculated using the KKR-ASA
CPA method \cite{pps3}. We have used the CPA rather than a rigid-band
model because CPA has been found to reliably describe the effects of
disorder in metallic alloys. We parametrized the exchange-correlation
potential as suggested by Perdew-Wang \cite{perdew1,perdew2} within the generalized
gradient approximation. The Brillouin zone (BZ) integration was carried
out using \protect\( 1215\protect \)  {\bf k}-points in
the irreducible part of the BZ. For DOS calculations, we added a small
imaginary component of \protect\( 2\protect \) \protect\( mRy\protect \)
to the energy and used \protect\( 3887\protect \)}
{\bf k}-\textmd{points in the irreducible
part of the BZ. The lattice constants for \protect\( MgB_{2}\protect \)
and \protect\( TaB_{2}\protect \) were taken from experiments while for
other compositions we used the Vegard's law. The Wigner- Seitz radii for
\protect\( Mg\protect \) and \protect\( Ta\protect \) were slightly larger
than that of \protect\( B\protect \). The sphere overlap, which is crucial
in ASA, was less than \protect\( 10\protect \)\% and the maximum
\protect\( l\protect \) used was \protect\( l_{max}\protect \) =
\protect\( 3\protect \).}}

\paragraph{\textmd{The electron-phonon coupling constant \protect\(
\lambda \protect \) was calculated using Gaspari-Gyorffy \cite{gaspari}
formalism with the charge self-consistent potentials of \protect\(
Mg\protect \)\protect\( _{1-x}\protect \)\protect\( Ta\protect
\)\protect\( _{x}\protect \)\protect\( B_{2}\protect \) obtained with the
KKR-ASA CPA method. Subsequently, the variation of \protect\(
T_{c}\protect \) as a function of \protect\( Ta\protect \) concentration
was calculated using Allen-Dynes equation \cite{allen}. The average values
of phonon frequencies \protect\( \omega _{ln}\protect \) for \protect\(
TaB_{2}\protect \) and \protect\( MgB_{2}\protect \) were taken from Refs.
\cite{rosner} and \cite{kortus} respectively. For intermediate concentrations, we
took \protect\( \omega _{ln}\protect \) to be the concentration-weighted
average of \protect\( MgB_{2}\protect \) and \protect\( TaB_{2}\protect
\). }}

\begin{figure}
\vspace{4cm}
\centering
\psfig{file=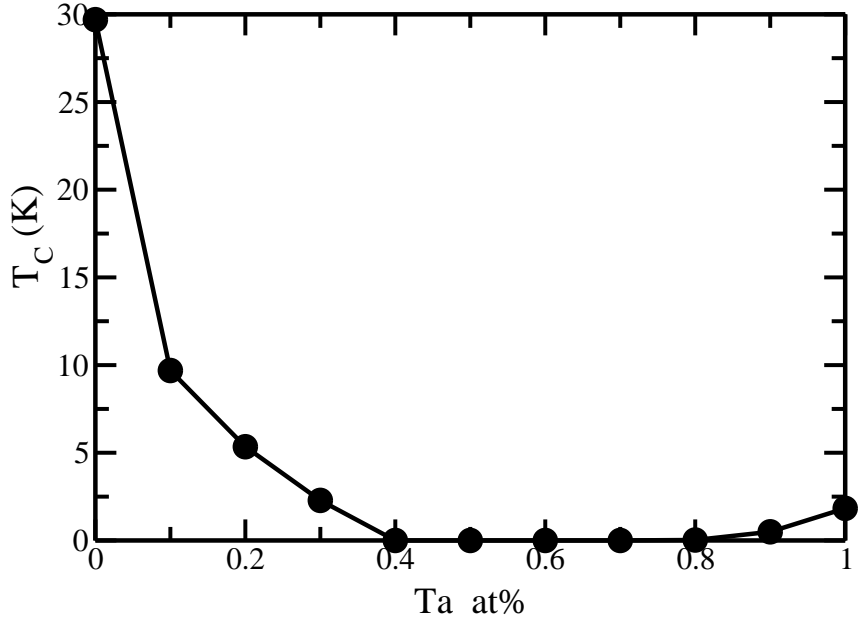,height=10cm,angle=-90}
\vspace{0.3cm}

\caption{The calculated variation of superconducting transition temperature
\protect\( T_{c}\protect \) with \protect\( Ta\protect \) concentration in
\protect\( Mg_{1-x\, }Ta_{x}\, B_{2}\protect \), with \protect\( \mu
^{*}=0.09\protect \).}

\end{figure}

\paragraph{\textmd{The main results of our calculations are shown in Fig.
1, where we have plotted the variation in \protect\( T_{c}\protect \) of
\protect\( Mg_{1-x}Ta_{x}B_{2}\protect \) alloys as a function of
\protect\( Ta\protect \) concentration. The calculations were carried out
as described earlier with the same value of \protect\( \mu
^{*}=0.09\protect \) for all the concentrations. The \protect\(
T_{c}\protect \) for \protect\( MgB_{2}\protect \) is equal to \protect\(
\sim 30\, K,\protect \) which is consistent with the results of other
works \cite{kortus} with similar approximations. The corresponding
\protect\( \lambda \protect \) is equal to \protect\( 0.73.\protect \) As
a function of \protect\( Ta\protect \) concentration we find that the
\protect\( T_{c}\protect \) decreases rapidly and goes to zero at around
\protect\( 40\, at\%\protect \) of \protect\( Ta\protect \) as shown in
Fig. 1. The \protect\( T_{c}\protect \) remains essentially zero until
around \protect\( 75-80\, at\%\protect \) of \protect\( Ta\protect \),
thereafter rising to \protect\( 1.8\, K\protect \) for \protect\(
TaB_{2}\protect \) with \protect\( x=1.\protect \) The corresponding
\protect\( \lambda \protect \) for \protect\( TaB_{2}\protect \) is equal
to \protect\( 0.38.\protect \).}}

\begin{figure}
\vspace{4cm}
\psfig{file=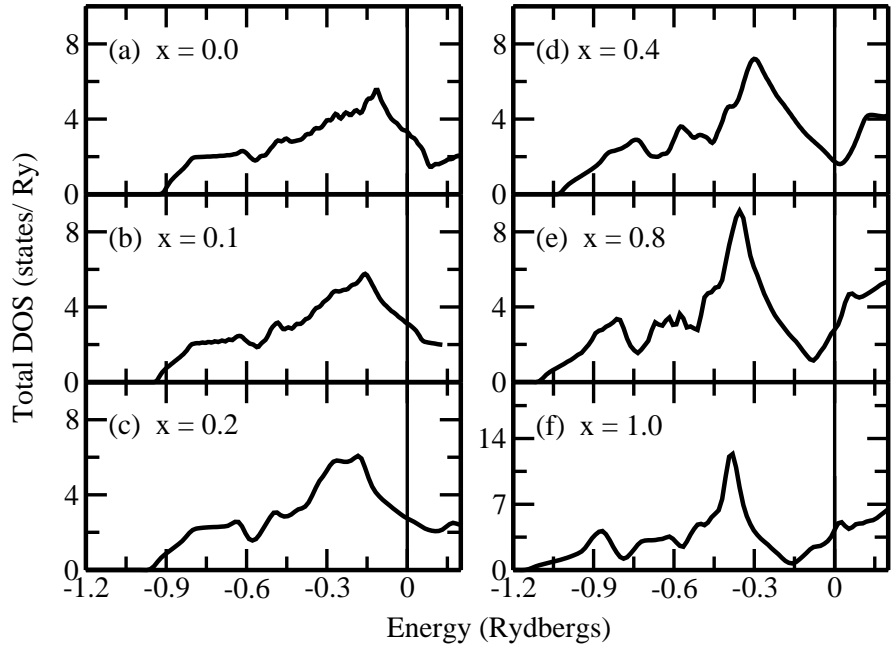,height=10cm,angle=-90}
\vspace{0.3cm}
\caption{The total density of states for \protect\( Mg_{1-x}Ta_{x}B_{2}\protect \) 
as a function of Ta concentration $x$, calculated
with the charge self-consistent KKR-ASA CPA method, as described in the text. Note that in (f),
the vertical scale
is different from the rest of the figures. The vertical line  denotes the Fermi energy.}
\end{figure}
\begin{table}
\vspace{0.3cm}
{\centering \begin{tabular}{|c|c|c|c|c|}
\hline 
&
\( a \) \( (a. \) \( u.) \)&
\( c/a \)&
\( N(E_{F}) \)&
\( B \) \( _{p} \)\\
\hline 
\hline 
\( MgB_{2} \)&
5.834&
1.141&
3.331&
2.834\\
\hline 
\( Mg_{0.8}Ta_{0.2}B_{2} \) &
5.833&
1.124&
3.116&
1.785\\
\hline 
\( Mg_{0.6}Ta_{0.4}B_{2} \)&
5.831&
1.106&
0.466&
0.616\\
\hline 
\( Mg_{0.4}Ta_{0.6}B_{2} \)&
5.830&
1.088&
1.734&
0.663\\
\hline 
\( Mg_{0.2} \)\( Ta_{0.8}B_{2} \)&
5.829&
1.070&
2.791&
0.833\\
\hline 
\( TaB_{2} \)&
5.828&
1.052&
4.980&
1.455\\
\hline 
\end{tabular}\par}
\vspace{0.3cm}

\caption{The calculated lattice parameter \protect\( a\protect \)
\protect\( (a.u.)\protect \),  \protect\( c/a\protect \), total DOS at the 
 Fermi energy 
(states/Ry/atom) and \protect\( B\protect \) $p$  partial DOS (states/Ry/spin).
The lattice parameters for intermediate concentrations were determined using Vegard's law.}
\end{table}
\paragraph{\textmd{
In order to get further insight into the
variation of \protect\( T_{c}\protect \) as a function of \protect\(
Ta\protect \) concentration in \protect\( Mg_{1-x}Ta_{x}B_{2}\protect \)
 alloys, we have analyzed the total DOS as well as the contribution of
$B$    \protect\( p\protect \)-electrons to the total DOS. 
In Figs. \protect\( 2(a)-(f)\protect \), we show the
total DOS of \protect\( Mg_{1-x}Ta_{x}B_{2}\protect \) alloys for
\protect\( Ta\protect \) concentration ranging from \protect\( x=0\protect
\) to \protect\( x=1,\protect \) calculated using the KKR-ASA CPA method
as described earlier. The DOS for \protect\( MgB_{2}\protect \) and
\protect\( TaB_{2}\protect \), as shown in Figs. \protect\( 2(a)\protect \)
and \protect\( 2(f)\protect \) are similar to previous calculations
\cite{kortus,pps2}. With small addition of \protect\( Ta\protect \) the
Fermi energy, \protect\( E_{F}\protect \), moves outward to accommodate the
additional electrons resulting in a decrease in the total DOS at \protect\(
E_{F}\protect \) for \protect\( Mg_{1-x}Ta_{x}B_{2}\protect \) alloys,
leading to a decrease in \protect\( T_{c}.\protect \) Further addition of
\protect\( Ta\protect \) allows \protect\( E_{F}\protect \) to be pinned
in a valley around \protect\( 50-60\, at\, \%\protect \) of \protect\(
Ta\protect \), resulting in a decreased donor ability of the metallic
plane (\protect\( Mg-Ta\protect \)) with the increase in the number of
\protect\( d\protect \) electrons. Thus the overall contribution of
\protect\( B\protect \) electronic states decreases as the concentration
of \protect\( Ta\protect \) is increased up to around \protect\( 80\,
at\%\protect \) of \protect\( Ta.\protect \) Starting with around
\protect\( 80\, at\%\protect \) of \protect\( Ta\protect \) the \protect\(
B\protect \) contribution rises and so does the \protect\( T_{c}\protect
\) as shown in Fig. 1. To substantiate these qualitative observations we
have listed in Table I the total DOS and the \protect\( B\, p\protect \)
DOS at \protect\( E_{F}.\protect \)
In addition, in Fig. 3, we show the \protect\( B\, p_{x}\protect
\) and \protect\( p_{z}\protect \) partial DOS as a function of
concentration.
It is clear from  Table I and Fig. 3 that the
\protect\( B\, p\protect \) DOS is responsible for the variation in
\protect\( T_{c}\protect \) of \protect\( Mg_{1-x}Ta_{x}B_{2}\protect \)
alloys, and 
 the possible loss of superconductivity in these alloys
at around \protect\( 60\, at\%\protect \)
of \protect\( Ta\protect \) can be attributed to a very small \protect\(
B\, p_{x,y}\protect \) DOS at \protect\( E_{F}.\protect \)  
 In Table I, we have also listed the
lattice constants used in our calculations. 
}}

\begin{figure}
\vspace{4cm}
\centering
\psfig{file=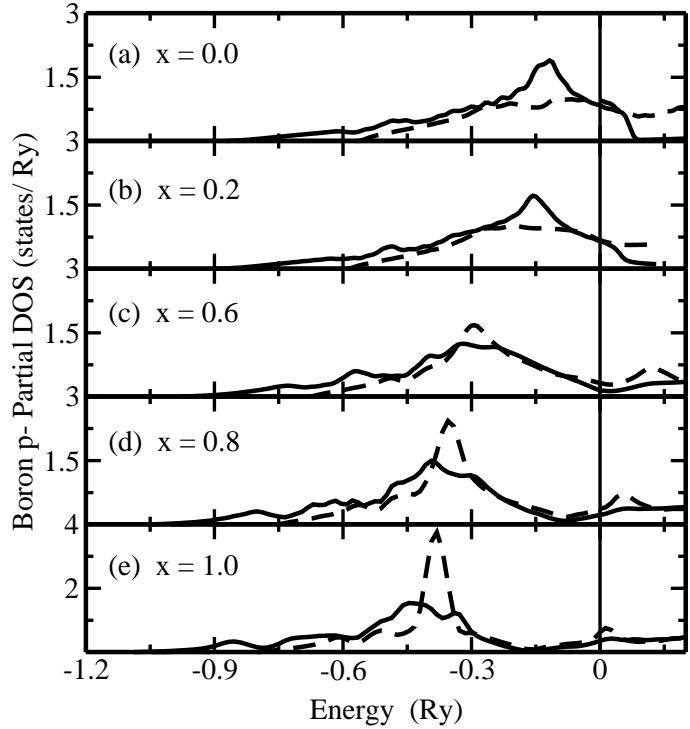,height=10cm,angle=-90}
\vspace{0.3cm}

\caption{\protect\( B\protect \) \protect\( p\protect \) partial
DOS showing \protect\( p_{x}\protect \) (full line) and \protect\(
p_{z}\protect \) (dashed line) for \protect\( Mg_{1-x\, }Ta_{x\,
}B_{2}\protect \). The vertical line denotes the Fermi energy.
Note that for \protect\( x=1\protect \) the vertical scale is different from the rest of the
plots.}
\end{figure}

\paragraph{\textmd{In conclusion, we have studied the variation of
\protect\( T_{c}\protect \) in \protect\( Mg_{1-x}Ta_{x}B_{2}\protect \)
alloys as a function of \protect\( Ta\protect \) concentration. We have
used the KKR-ASA CPA method for taking into account the effects of
disorder and Gaspari-Georffy formalism for calculating the electron-phonon
coupling constant \protect\( \lambda \protect \). The \protect\(
T_{c}\protect \) is then calculated using the Allen-Dynes equation. Our
results show that the \protect\( T_{c}\protect \) decreases with the
addition of \protect\( Ta\protect \) for upto \protect\( 40\,
at\%\protect \), thereafter remains essentially zero up to \protect\( 80\,
at\%\protect \) of \protect\( Ta.\protect \) We find \protect\(
TaB_{2}\protect \) to be superconducting, albeit at a low temperature. We
have also shown that the variation in \protect\( T_{c}\protect \) is
mostly dictated by the changes in the \protect\( B\, p\protect \)
density of states.}}

\paragraph{\textmd{We would like to thank Dr. I. A. Abrikosov for
providing a copy of his locally self-consistent Green's function code.}}


\begin{thebibliography}{10}

\bibitem{nag}\( \)J. Nagamatsu, N. Nakagawa, T. Muranaka, Y. Zenitani, and
J. Akimitsu, Nature, 410, 63 (2001).

\bibitem{budko1} S. L. Bud'ko, C. Petrovic, G. Lapertot, C. E. Cunningham,
P. C. Canfield, M. H. Jung, A. H. Lacerda, cond-mat/ 0102413.

\bibitem{kang}W. N. Kang, C. U. Jung, Kijoon H. P. Kim, Min-Seok Park, S.
Y. Lee, Hyeong-Jin Kim, Eun-Mi Choi, Kyung Hee Kim, Mun-Seog Kim, and
Sung-Ik Lee, cond-mat/ 0102313.

\bibitem{buzea}Cristina Buzea and Tsutomu Yamashita, cond-mat/ 0108265;
and references therein.

\bibitem{ann}M. An and W. E. Pickett, cond-mat 0102391.

\bibitem{hirsch1}J. E. Hirsch, cond-mat/0102115.

\bibitem{belash}K. D. Belashchenko, M. van Schilfgaarde, and V. P.
Antropov, cond-mat/ 0102290.

\bibitem{imada} M. Imada, cond-mat/0103006.

\bibitem{hirsch2}J. E. Hirsch and F. Marsiglio, cond-mat/ 0102479.

\bibitem{budko2} S. L. Bud'ko, G. Lapertot, C. Petrovic, C. E. Cunningham,
N. Anderson, and P. C. Canfield, cond-mat/ 0101463.

\bibitem{kong}Y. Kong, O. V. Dolgov, O. Jepsen, and O. K. Andersen,
cond-mat/ 0102499.

\bibitem{pps1}Prabhakar P. Singh, cond-mat/ 0104580.

\bibitem{felner}Israel Felner, cond-mat/ 0102508.

\bibitem{slusky}J. S. Slusky, N. Rogado, K. A. Regan, M. A. Hayward, P.
Khalifah, T. He, K. Inumaru, S. Loureiro, M. K. Haas, H. W. Zandbergen,
and R. J. Cava, cond-mat/ 0102262.

\bibitem{kazakov}S. M. Kazakov, M. Angst, and J. Karpinski, cond-mat/
0103350.

\bibitem{suzuki}S. Suzuki, S. Higai, and K. Nakao, cond-mat/ 0102484.

\bibitem{zhao}Y. G. Zhao, X. P. Xang, P. Y. Qiao, H. T. Zhang, S. L. Jia,
B. S. Cao, M. H. Zhu, Z. H. Han, X. L. Wang and B. L. Gu,
cond-mat/0104063.

\bibitem{morotomo}Y. Morotomo and Sh Xu, cond-mat/0104568.

\bibitem{kac}D. Kaczorowski, A. J. Zaleski, O. J. Zogal, and J. Klamut,
cond-mat/0103571.

\bibitem{laya}L. Layarovska and E. Layarovski, J. Less Common Met 67, 249
(1979).
\bibitem{faulkner}J. S. Faulkner, Prog. Mat. Sci 27, 1 (1982); and
references therein.



\bibitem{pps3}Prabhakar P. Singh and A. Gonis, Phys. Rev. B 49, 1642
(1994).

\bibitem{perdew1}J. P. Perdew and Y. Wang, Phys. Rev. B 45, 13244 (1992);

\bibitem{perdew2}J. P. Perdew, K. Burke and M. Ernzerhof, Phys. Rev. Lett.
77, 3865 (1996).

\bibitem{gaspari}G. D. Gaspari and B. L. Gyorrfy, Phys. Rev. Lett. 28, 801
(1972).

\bibitem{allen}P. B. Allen and R. C. Dynes, Phys. Rev. B 12, 905 (1975).
\bibitem{rosner}H. Rosner and W. E. Pickett, cond-mat/0100092 (2001).

\bibitem{kortus}J. Kortus, I. I. Mazin, K. D. Belashchenko, V. P.
Antropov, L. L. Boyer, cond-mat/0101446.

\bibitem{pps2}Prabhakar P. Singh, cond-mat/ 0104563.

\end{thebibliography}
\end{document}